# Characterizing Fungal Infections in the *All of Us* Research Program

*Md Fantacher Islam, MS[1], Emeral Norzagaray, BS[2], Corneliu Antonescu, MD[3], Vignesh Subbian, PhD[1]

[1]College of Engineering, The University of Arizona, Tucson, Arizona, United States

[2]College of Agriculture, Life & Environmental Sciences, The University of Arizona, Tucson, Arizona, United States

[3]College of Medicine Phoenix, The University of Arizona, Phoenix, Arizona, United States

*Corresponding Author: Md Fantacher Islam

E-mail: fantacher@arizona.edu

Keywords: fungal diseases, fungal pathogens, coccidioidomycosis, aspergillosis, histoplasmosis

**Abstract**

Fungal infections, such as coccidioidomycosis, aspergillosis, and histoplasmosis, represent a growing public health concern in the United States. However, the actual burden of these infections is often underestimated by traditional surveillance methods. Therefore, this study characterizes these infections within the *All of Us* Research Program and evaluates the quality of related clinical data. We constructed three cohorts, coccidioidomycosis (n = 1,171), aspergillosis (n = 687), and histoplasmosis (n = 347), among 393,601 participants with electronic health record data. We analyzed geographic and sociodemographic distributions, disease reporting, and host factors, and performed a data quality assessment on 10 key laboratory biomarkers and organism specific tests within 90 days of diagnosis. Our analysis confirmed known epidemiological patterns, including the geographic distributions of coccidioidomycosis in the Southwest and histoplasmosis in the Midwest. Aspergillosis is not routinely reportable, and 49.3% of histoplasmosis cases occurred in states without reporting mandates. Fungal infections disproportionately affected older adults and males relative to baseline, and 44.7% of aspergillosis cases were in immunocompromised patients. The data quality assessment showed high completeness for general hematology markers (e.g., hemoglobin > 70% in coccidioidomycosis and aspergillosis) but limited availability of organism-specific tests, such as *Coccidioides* serology (4.6% in coccidioidomycosis). Most coccidioidomycosis hemoglobin values (97.9%) were reported in g/L despite a g/dL label. In conclusion, the *All of Us* dataset is a valuable resource for characterizing fungal infections, but data quality issues related to completeness and conformance must be addressed to support robust clinical research.

## Introduction

Fungal infections are an understudied yet serious public health concern that causes millions of deaths globally, tens of millions of infections, and billions in United States (U.S.) healthcare costs.[1–3] In the early 2020s, hospitalizations from fungal diseases rose substantially, with mortality exceptionally high among patients who developed fungal infections during coronavirus disease 2019 (COVID-19) illness.[4] Among these fungal diseases, coccidioidomycosis (also known as Valley fever), histoplasmosis, and aspergillosis are among the most clinically and epidemiologically significant, contributing substantially to morbidity across diverse populations. The World Health Organization lists *Aspergillus fumigatus* as a critical priority fungal pathogen, and *Histoplasma* and *Coccidioides* species as high and medium priority pathogens.[5] These mycoses have demonstrated rising incidence in recent years, driven in part by environmental changes, shifts in population demographics, and expanding susceptible host populations.[6]

Coccidioidomycosis is an infection from inhaling *Coccidioides* spores in desert soil that is often minimally symptomatic or self-limited but can progress to a lung infection or spread throughout the body. This disease disseminates more often in Black and Filipino patients, during pregnancy, and in people with weakened immune systems.[7,8] Aspergillosis is caused by *Aspergillus* molds and can range from allergic bronchopulmonary aspergillosis to a severe invasive disease, which is particularly dangerous for patients with weakened immune systems.[9] The invasive forms of this infection have a 50% fatality rate and can reach 90% in vulnerable groups, while conditions related to COVID-19 amplify the risk of pulmonary aspergillosis and have a mortality above 40%.[10,11] Finally, histoplasmosis is another form of fungal infection, also acquired by inhaling *Histoplasma capsulatum* spores from soil contaminated with bird or bat droppings, and its symptoms are often vague, which can delay its diagnosis. The disease was once thought to be limited to the Ohio and Mississippi River valleys, but it is now found in most U.S. states. In 2019, around 1,124 cases were reported, and over half of these patients were hospitalized, and approximately 5% died.[12–14] It occurs mainly in people with impaired cellular immunity, including people with human immunodeficiency virus (HIV).[15]

Risk factors for these infections differ: coccidioidomycosis and histoplasmosis are associated with soil exposure in endemic areas, whereas invasive aspergillosis is primarily linked to weakened immune systems.[7,9,15] Social factors, such as crowded living conditions and limited access to healthcare, can also increase exposure and delay diagnosis.[16] Studies also suggest that these social and environmental vulnerabilities contribute more to racial and ethnic disparities of fungal infections than genetic factors.[17] However, national surveillance by the Centers for Disease Control and Prevention (CDC) provides only a limited view of the true burden. Coccidioidomycosis is nationally notifiable, but reporting depends on state rules. In 2019, coccidioidomycosis was reportable in 26 states, and histoplasmosis in 13 states, whereas aspergillosis is not routinely reportable.[14,18] According to CDC surveillance reports from 2019 to 2021, nearly 60,000 coccidioidomycosis cases and thousands more for histoplasmosis were reported, with rates hitting four times higher among American Indian or Alaska Native people and twice as high for Hispanic individuals compared to White individuals who are not Hispanic.[19] Yet modeling suggests that true symptomatic coccidioidomycosis incidence may be 10 to 18 times higher, corresponding to about 23,000 hospitalizations and about 900 deaths in a single year.[20] These underestimates become more compounded by diagnostic challenges, as fungal pneumonias frequently mimic bacterial or viral diseases, leading to delays and inappropriate antibiotic use. A study using real-world data from Oracle electronic health records found high rates of histoplasmosis in states that do not report the disease.[21] CDC investigators suggested that linked data resources, such as the *All of Us* Research Hub, could help study disparities in fungal diseases.[22] Most EHR data are collected for clinical care rather than for research, so their quality should be assessed before use in research. In *All of Us*, harmonized data quality frameworks have been implemented across diseases.[23,24] However, the suitability of *All of Us* data for fungal disease research remains unassessed.

Therefore, the goals of this study are (a) to characterize the demographic, geographic, and socioeconomic burden of coccidioidomycosis, histoplasmosis, and aspergillosis, including host factors and cases outside disease reporting, within the *All of Us* Research Program cohort and (b) to conduct a data quality assessment related to the fitness-for-use of key clinical biomarkers for mycology research in the *All of Us* Program.

## Materials and methods

### The *All of Us* Research Program

This study used data from the National Institutes of Health (NIH) *All of Us* Research Program, which was launched in 2018 to enroll over one million diverse U.S. participants to accelerate health research and improve public health outcomes.[25,26] *All of Us* Research Program collects data through electronic health records (EHRs), participant surveys, physical measurements, and genomic analyses.[26] We used version 8 curated data repository, which includes 633,547 participants, 393,601 of whom have linked EHR data obtained from a nationwide network of healthcare organizations. This study did not require approval from the Institutional Review

Board (IRB) because *All of Us* IRB determined that the research data, including Controlled Tier data used in this research, are de-identified and do not constitute human subject research. All direct identifiers, including free-text fields, had been removed from the data, and all dates had been systematically shifted to reduce the risk of re-identification, in accordance with *All of Us* privacy safeguards. Data analyses were conducted within the secure *All of Us* Researcher Workbench under the Data User Code of Conduct, which explicitly prohibits any attempt to re-identify participants.[27] All data analyses were performed using Python 3.12.

**Study Population and Case Identification**

Our inclusion criteria were set to include participants with linked EHR data who had at least one diagnosis record of fungal infections in the *All of Us* Research Program. Cohorts were identified using specific codes mapped from Systematized Nomenclature of Medicine Clinical Terms (SNOMED-CT) as well as the Observational Medical Outcomes Partnership (OMOP) Common Data Model. A complete list of codes and their corresponding OMOP concept IDs used for coccidioidomycosis cohort identification is provided in **Table 1**. Similar approaches were applied for identifying cohorts of aspergillosis and histoplasmosis in Supplementary Tables S1 and S2.

**Data Extraction and Data Quality Assessment**

For the three fungal infection cohorts, we extracted demographic, geographic, socioeconomic, and biomarker data. Demographic variables included age at first diagnosis, sex assigned at birth, and self-reported race, ethnicity, education, and income reported by participants in survey. Geographic information was derived from participants' state of residence. To ensure the reliability of laboratory data (e.g., erythrocyte sedimentation rate, C-reactive protein), we conducted a formal data quality assessment across three domains: completeness, conformance, and concordance, following the harmonized data quality framework used in *All of Us*.[23,24] In this study, completeness reflects the availability of laboratory data, defined as the percentage of patients in each cohort with at least one measurement. Conformance assesses the consistency of units of measure across sites, and concordance evaluates the agreement of values through correlations between related markers. Before calculating correlations, we removed values outside adult physiologic ranges, including hemoglobin 3 to 25 g/dL and CRP 0 to 500 mg/L, and converted hemoglobin values from 30 to 250 g/L to g/dL.

**Laboratory Data Related to Fungal Infections**

We extracted and analyzed clinically relevant laboratory biomarkers routinely captured in real-world data.[28] C-reactive Protein (CRP) and Erythrocyte Sedimentation Rate (ESR) were included to capture nonspecific inflammation common in fungal infections.[29] Procalcitonin was included to help differentiate bacterial coinfections, while 1,3-β-D-glucan helped in validating a panfungal biomarker.[30] Eosinophil counts were included for their utility in phenotype differentiation, particularly for allergic bronchopulmonary aspergillosis.[7,9] Lactate dehydrogenase (LDH) and alkaline phosphatase (ALP) were used to assess disseminated disease and organ involvement, especially in histoplasmosis.[15] Finally, hemoglobin and hematocrit were included as physiologic anchors for internal quality control. To ensure temporal relevance, laboratory data were restricted to a window within 90 days from the initial diagnosis date, consistent with established infectious disease conventions.[31] Because diagnosis codes alone do not indicate whether laboratory testing confirmed an infection, we also assessed organism-specific tests used to diagnose these infections. We included *Coccidioides* serology or antigen,[7] *Histoplasma* antigen or serology,[15]

*Aspergillus* galactomannan or antibody,[9] and fungal culture or stain. We identified these tests by concept name within the same 90-day window.

**Prevalence and disease reporting**

We analyzed first diagnoses documented from 2013 to 2023 among adult participants in the U.S. We calculated period prevalence per 100,000 participants with EHR data (n = 393,413) and exact binomial 95% confidence intervals (CIs), omitting states with 20 or fewer cases per program dissemination guidelines. We also calculated annualized prevalence by dividing the period prevalence by the 11 years of the study period. We classified cases based on whether the condition was reportable in the participant's state in 2019.[14]

**Host factors**

We selected host factors from consensus definitions of invasive fungal disease and established risk factors for each infection.[7,9,32] We identified these conditions by searching condition concept names for terms matching each host factor in records with onsets on or before the index date, which was defined as the first diagnosis of any of the three infections. These include HIV infection, solid organ or hematopoietic cell transplantation, hematologic malignancy, other malignancies, neutropenia within 90 days before the index date, diabetes mellitus, chronic obstructive pulmonary disease (COPD) or bronchiectasis, and asthma. We also identified systemic corticosteroids administered at least twice within 60 days before the index date, along with other immunosuppressive medications within 90 days before the index date.[32] Because doses were not reliably recorded, two or more administrations were used to approximate prolonged corticosteroid use. Additionally, we recorded antifungal drugs, such as azoles, echinocandins, and amphotericin B,[9] if administered within 90 days before or after the index date. A participant was considered immunocompromised if they had HIV, transplantation, hematologic malignancy, neutropenia, systemic corticosteroids, or other immunosuppressive drugs.

## Results

### Prevalence and disease reporting

There was a total of 633,547 participants in the *All of Us* program's version 8, including 393,601 participants with linked EHR data. The fungal infection cohorts included 1,171 individuals with coccidioidomycosis, 687 with aspergillosis, and 347 with histoplasmosis. The geographic distribution of these cohorts is presented in **Figure 1.**

Nearly all coccidioidomycosis cases were concentrated in the southwestern U.S., with the highest prevalence in Arizona (1,810). Aspergillosis cases were widely distributed across the U.S., with the highest prevalence in Arizona (272.6), Massachusetts (173.2), Florida (161.9), and California (161.6). In contrast, histoplasmosis prevalence was highest in the Midwest, in Michigan (142.6) and Illinois (83.6). Across the U.S., the prevalence was 284.9 for coccidioidomycosis, 145.4 for aspergillosis, and 57.7 for histoplasmosis (**Table 2**), with annualized prevalence of 25.9, 13.2, and 5.2, respectively. Almost all coccidioidomycosis cases (1,098) lived in states where the disease was reportable. However, aspergillosis is not routinely reportable, and 112 (49.3%) histoplasmosis cases lived in states without a reporting requirement, including Missouri and Iowa.

**Sociodemographic Characteristics**

The sociodemographic characteristics of the three cohorts are presented in Table 3. Compared with all participants with EHR data, in which 38.1% were men, men were disproportionately represented in the fungal infection cohorts, accounting for 541 (46.2%) coccidioidomycosis, 316 (46.0%) aspergillosis, and 158 (45.5%) histoplasmosis cases. Nevertheless, women accounted for a slight majority of cases within each cohort. Most individuals with fungal infections were older, with the age group 65 and above representing the largest age group across all three cohorts (40.2%, 49.5%, and 58.8%, compared with 34.4% of all participants with EHR data).

Racial and ethnic distributions within the cohorts also differed from those of all participants with EHR data. The coccidioidomycosis cohort had a higher proportion of Hispanic participants, with 277 cases (23.7%) compared with 18.2% overall. In contrast, White participants were more common in the aspergillosis (63.3%) and histoplasmosis (76.4%) cohorts, and Black participants were less common across all three cohorts. Socioeconomic indicators showed similar patterns: the coccidioidomycosis cohort had more participants with annual household incomes below $25,000 and fewer with an advanced degree.

**Host factors**

Host factors varied across the three fungal cohorts (**Table 4**). The aspergillosis cohort had the highest proportion of immunocompromising conditions, with nearly half of cases (307) immunocompromised. Compared with the coccidioidomycosis cohort, the aspergillosis cohort had higher proportions of transplantation (18.5% vs 6.7%), hematologic malignancies (12.2% vs 3.6%), and systemic corticosteroid use (25.8% vs 9.6%). Chronic lung disease, including COPD or bronchiectasis and asthma, were also more common. In contrast, HIV was most common in the histoplasmosis cohort (28), whereas diabetes was most common in the coccidioidomycosis cohort (460). Antifungal treatment also differed across cohorts, with 307 (44.7%) aspergillosis and 356 (30.4%) coccidioidomycosis cases, but only 62 (17.9%) histoplasmosis cases.

**Data Quality Assessment**

Assessment of data quality for laboratory tests and biomarkers related to fungal infections revealed variability across three key domains: data completeness, unit conformance, and measurement concordance.

**Completeness**

We assessed data based on the proportion of participants in each cohort who had at least one measurement within 90 days of diagnosis (**Table 5)**. In the analysis cohort, completeness was high for hemoglobin and hematocrit in the coccidioidomycosis (75.0%) and aspergillosis (73.1% to 73.5%) cohorts and lower in the histoplasmosis cohort (54.8% to 56.2%). However, inflammatory markers had low completeness across all cohorts, with CRP, ESR, and LDH each recorded for 26% or fewer of participants. Procalcitonin and 1,3-β-D-glucan were measured in about 10% of the aspergillosis cohort and in 20 or fewer participants in the other cohorts. We also found that organism-specific tests were rarely recorded. *Coccidioides* serology or antigen was recorded for only 4.6% of the coccidioidomycosis cohort, and *Histoplasma* antigen or serology for 12.1% of the histoplasmosis cohort, whereas *Aspergillus* galactomannan or antibody was recorded for 37.1% of the aspergillosis cohort.

### Conformance

CRP and hemoglobin showed high unit conformance in the coccidioidomycosis cohort, with a single text unit (mg/L and g/dL) accounting for over 95% of records. However, unit conformance was lower for these markers in the aspergillosis cohort (56.7% and 79.2%). In the coccidioidomycosis cohort, 94.5% of absolute eosinophil records and 93.3% of ESR records had no matching unit concept. We also identified a different conformance issue for alkaline phosphatase and LDH, where the most common unit in the coccidioidomycosis cohort was the numerical code 258947003. This code is the SNOMED CT concept for the unit U/L, which conforms to a standard vocabulary but is not readily human-readable. Finally, we found a hidden issue with hemoglobin that unit checks alone would not detect. In the coccidioidomycosis cohort, 97.4% of records were labeled g/dL (**Table 6**), yet 97.9% of values were between 30 and 250, the range expected for g/L.

### Concordance

We calculated measurement concordance using bivariate Spearman correlations among biomarkers for each of the three fungal cohorts. We observed a strong positive correlation among hematologic markers in the aspergillosis cohort. Hemoglobin and hematocrit were very strongly correlated across all cohorts (r = 0.97 to 0.98) (**Figure 2)**. Correlations among inflammatory markers were weaker and varied by cohort. CRP and ESR were moderately correlated across all three cohorts (r = 0.49, 0.44, and 0.44), whereas LDH could not be paired with CRP or ESR in the histoplasmosis cohort because 20 or fewer participants had both measurements.

## Discussion

Our analysis highlights geographic patterns for coccidioidomycosis, aspergillosis, and histoplasmosis-related infections in the *All of Us* Research Program, which align with known epidemiological knowledge of these mycoses in the United States. Coccidioidomycosis cases were concentrated in Arizona and California, which accounted for the majority of infections and reflect the known endemicity of *Coccidioides* species in arid regions, where soil disturbance and dust inhalation facilitate transmission.[7,8] The prevalence in Arizona was about 20 times that in California, and this pattern aligns with 2019 CDC surveillance data, which report the highest incidence in Arizona.[14] Similarly, histoplasmosis prevalence was highest in Michigan and Illinois, which is consistent with the Ohio and Mississippi River valleys as hotspots for *H. capsulatum* due to bird and bat guano in moist soils.[13,15] In contrast, aspergillosis exhibited a more widespread distribution across states, highlighting its association with underlying immunosuppression rather than strict environmental factors. The highest prevalence of aspergillosis observed in Arizona aligns with a study using Oracle EHR Real-World Data.[33] However, because aspergillosis lacks a specific endemic area, its state distribution may also reflect where patients receive care. The estimated annual prevalence of histoplasmosis (5.2) and aspergillosis (13.2) was similar to recent estimates from the same database of 2.9 to 6.5 and 15.3 per 100,000 person-years, respectively.[21,33]

Demographic characteristics revealed vulnerabilities among older adults, males, and certain ethnic groups across the fungal cohorts. Individuals aged 65 and older made up a larger proportion of each cohort than of all participants with EHR data, which suggests an age-related risk factor. Men were also more common in the cohorts than among all participants with EHR data. The coccidioidomycosis cohort had more Hispanic participants, and the histoplasmosis cohort had more White participants, following the same trend observed in CDC surveillance data for 2019.[14] Additionally, the coccidioidomycosis cohort showed socioeconomic disparities, with lower education and income levels than all participants with EHR data. These findings highlight studies linking lower socioeconomic status to increased environmental exposure and barriers to early diagnosis, emphasizing the need for equitable access to screening in underserved communities.[16]

Host factors differed across the three fungal cohorts and aligned with the known risk groups for each infection. The aspergillosis cohort had the highest prevalence of immunocompromising conditions and chronic lung disease, consistent with the established risk of aspergillosis among patients with transplantation, hematologic malignancy, corticosteroid use, and chronic obstructive pulmonary disease.[2,9] In contrast, diabetes was most common in the coccidioidomycosis cohort, and diabetes is a known risk factor for symptomatic coccidioidomycosis.[7] HIV was most common in the histoplasmosis cohort, reflecting the higher risk of disseminated histoplasmosis among people with acquired immunodeficiency syndrome (AIDS). Fewer histoplasmosis cases received antifungal treatment, which may reflect that many cases of acute pulmonary histoplasmosis resolve without treatment.[15] However, treatment received outside participating health systems could not be captured.

The data quality assessment highlights both strengths and challenges of using *All of Us* EHR data for secondary research on fungal diseases. Data completeness was high for hematology markers, such as hemoglobin and hematocrit, in the coccidioidomycosis and aspergillosis cohorts, supporting population-level analyses of anemia or inflammation associated with these infections. However, the low capture of organism-specific tests may reflect testing outside participating health systems or recording without standard codes, limiting the use of laboratory data to confirm diagnosis codes. Unit conformance was high for markers such as C-reactive protein and hemoglobin in the coccidioidomycosis cohort, but other markers had issues. For example, ESR and absolute eosinophil counts had the most unmatched units, likely due to data collection from diverse EHR systems. In addition, hemoglobin values labeled g/dL but recorded in g/L suggest mapping errors at the site level that would be missed without checking value ranges.

Concordance analyses revealed strong correlations between related markers, such as hemoglobin and hematocrit, while correlations between inflammatory markers were weaker and some could not be estimated in the histoplasmosis cohort because of few paired measurements.

Limitations of this study include the potential for misclassification due to reliance on OMOP and SNOMED codes for case identification, which may lead to underestimation or overestimation of fungal infections without confirmatory serological or culture data. The relatively small cohort sizes, along with small counts for low frequency categories as per the *All of Us* data policy, limit statistical power and the ability to conduct granular subgroup analyses. *All of Us* participants are volunteers, so our prevalence estimates may not reflect the general U.S. population. Host factors were identified from diagnosis and drug records without chart review. Our evaluation was limited to selected biomarkers and organism-specific tests by concept name, possibly missing other fungus-specific markers in the dataset or clinical practice. Future research could mitigate these limitations by integrating *All of Us* genomic, environmental, or longitudinal data. Despite these constraints, the *All of Us* Research Program is a powerful resource for characterizing the epidemiology of these fungal diseases and confirming known demographic and geographic patterns.

## Conclusion

This study used the *All of Us* Research Program to characterize the epidemiology of coccidioidomycosis, aspergillosis, and histoplasmosis and to assess the quality of related laboratory data. Our findings confirm established endemic patterns and show that older adults and males were disproportionately affected. We also found that many aspergillosis and histoplasmosis cases occurred in states that do not report these diseases. Despite limitations such as small cohort sizes, the *All of Us* dataset is a valuable resource for epidemiological insights, with complete routine markers but limited capture of specialized biomarkers and organism-specific tests. Its laboratory values require value range checks and unit harmonization before use. Integrating the rich genomic, environmental, and longitudinal data within *All of Us* will help advance equitable mycology research to combat these understudied public health threats.

## Author contributions

Md Fantacher Islam (Data curation, Formal analysis, Investigation, Methodology, Software, Visualization, Writing - original draft), Emeral Norzagaray (Conceptualization, Data curation, Investigation, Visualization), Corneliu Antonescu (Supervision, Investigation, Validation, Writing - review and editing), Vignesh Subbian (Conceptualization, Project administration, Methodology, Supervision, Resources, Validation, Writing - review and editing)

## Data availability statement

This study used data from the *All of Us* Researcher Workbench (Controlled Tier Dataset version 8, 2025), and access requires a three-step authorization process that includes registration, completion of ethics training, and a Data Use Agreement at https://www.researchallofus.org.

## Figures

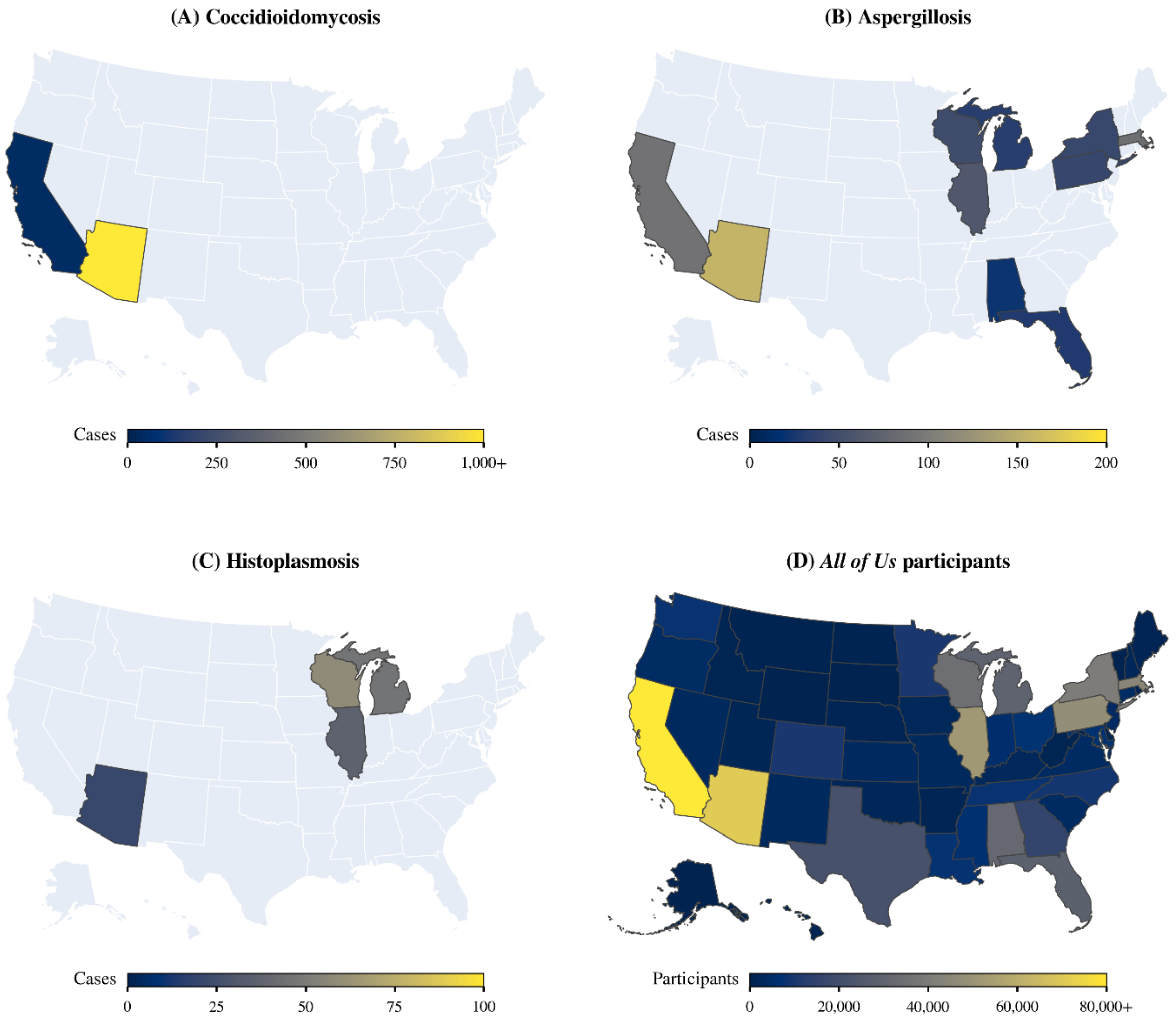

**Figure 1.** Geographic distribution of fungal infection cohorts in the *All of Us* Research Program. State-by-state distributions are presented for the (A) coccidioidomycosis cohort, (B) aspergillosis cohort, and (C) histoplasmosis cohort. Panel (D) shows the geographic distribution of the entire *All of Us* cohort for comparison. States with 20 or fewer cases are not displayed, in accordance with the *All of Us* Research Program data dissemination policy. Note: The map was created by the authors using publicly available cartographic boundary files from the U.S. Census Bureau (available at https://www.census.gov/geographies/mapping-files/time-series/geo/cartographic-

boundary.html) and the *All of Us* Research Program participant data. In accordance with the program policy, this work acknowledges the essential contributions of *All of Us* participants.

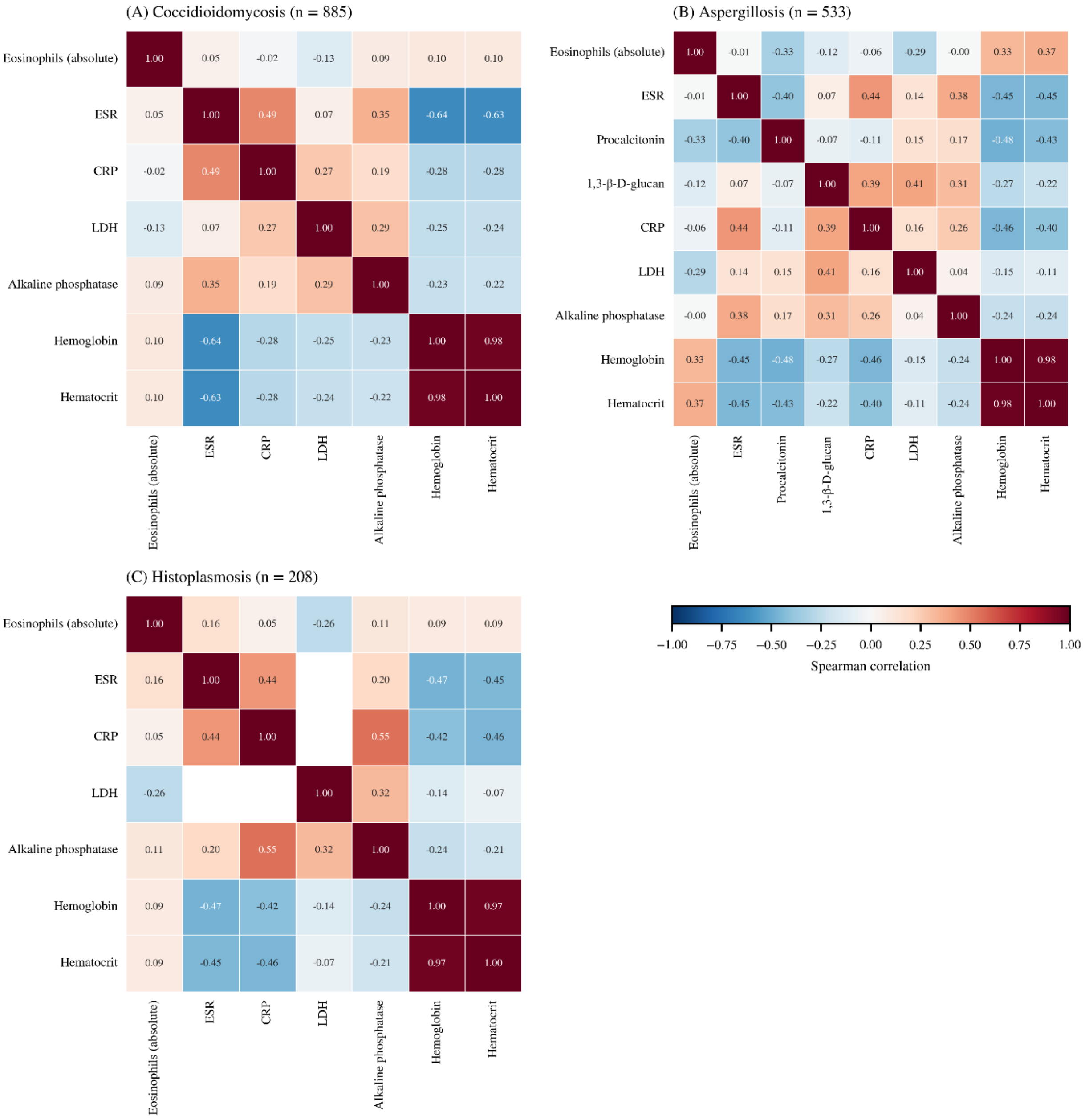

**Figure 2.** Correlogram of nine key laboratory biomarkers in the (A) coccidioidomycosis, (B) aspergillosis, and (C) histoplasmosis cohorts. Procalcitonin and 1,3-β-D-glucan are shown only for aspergillosis because 20 or fewer participants in the other cohorts had these tests. Correlations are computed using the median value for each participant after removing values outside physiologic ranges and converting hemoglobin from g/L to g/dL. We excluded eosinophil percentage, and pairs with 20 or fewer participants are left blank. Each panel title reports the number of participants with at least one value within these ranges for any of the 10 biomarkers. CRP, C-reactive protein. ESR, erythrocyte sedimentation rate. LDH, lactate dehydrogenase.

## Tables

**Table 1.** Concept IDs used to identify coccidioidomycosis in the electronic health records, *All of Us* Research Program.

| Description | OMOP Concept ID | SNOMED Code |
|---|---|---|
| Primary cutaneous coccidioidomycosis | 4197239 | 79949009 |
| Primary pulmonary coccidioidomycosis | 256909 | 88036000 |
| Pneumonia with coccidioidomycosis | 4110507 | 195904005 |
| Pulmonary coccidioidomycosis | 443741 | 417018008 |
| Acute pulmonary coccidioidomycosis | 4087776 | 187027001 |
| Primary extrapulmonary coccidioidomycosis | 438071 | 23247008 |
| Disseminated cutaneous coccidioidomycosis | 4300340 | 403110005 |
| Disseminated coccidioidomycosis | 4224517 | 85055004 |
| Coccidioidal meningitis | 435182 | 46303000 |
| Progressive coccidioidomycosis | 40493189 | 445880000 |
| Infection by *Coccidioides immitis* | 40484019 | 442543009 |
| Chronic pulmonary coccidioidomycosis | 260034 | 233615002 |
| Coccidioidomycosis | 437217 | 60826002 |
| Cutaneous coccidioidomycosis | 4082061 | 240727006 |

OMOP, Observational Medical Outcomes Partnership; SNOMED, Systematized Nomenclature of Medicine.

**Table 2.** Prevalence of first diagnoses of fungal infections by reporting status.

| Group | Cases, No. (%) | Prevalence per 100,000 (95% CI) |
|---|---|---|
| Coccidioidomycosis | 1,121 | 284.9 (268.5 to 302.1) |
| Aspergillosis | 572 | 145.4 (133.7 to 157.8) |
| Histoplasmosis | 227 | 57.7 (50.4 to 65.7) |
| *By reporting status of the state of residence in 2019* | | |
| Coccidioidomycosis, reportable | 1,098 (97.9) | 561.3 (528.7 to 595.4) |
| Coccidioidomycosis, not reportable | 23 (2.1) | 11.6 (7.4 to 17.4) |
| Histoplasmosis, reportable | 115 (50.7) | 86.6 (71.5 to 103.9) |
| Histoplasmosis, not reportable | 112 (49.3) | 43.0 (35.4 to 51.7) |
| Aspergillosis, not routinely reportable | 572 (100) | 145.4 (133.7 to 157.8) |
| *States with more than 20 cases* | | |
| Coccidioidomycosis | | |
| Arizona | 1,029 | 1,810 (1,702 to 1,922.9) |
| California | 45 | 90.9 (66.3 to 121.6) |
| Aspergillosis | | |
| Arizona | 155 | 272.6 (231.5 to 319) |
| Massachusetts | 60 | 173.2 (132.2 to 222.9) |
| Florida | 27 | 161.9 (106.7 to 235.5) |
| California | 80 | 161.6 (128.1 to 201) |
| New York | 36 | 127.9 (89.6 to 177) |
| Illinois | 45 | 125.5 (91.5 to 167.8) |
| Michigan | 24 | 114.1 (73.1 to 169.7) |
| Wisconsin | 28 | 104.8 (69.7 to 151.5) |
| Pennsylvania | 32 | 91.3 (62.5 to 128.9) |
| Histoplasmosis | | |
| Michigan | 30 | 142.6 (96.2 to 203.5) |
| Illinois | 30 | 83.6 (56.4 to 119.4) |

CI, confidence interval. The reporting status for the state of residence is as of 2019. States with 20 or fewer cases are not shown.

**Table 3.** Sociodemographic Characteristics of *All of Us* Fungal Infections Cohorts

| | Cohort | | | |
|---|---|---|---|---|
| **Category** | **Participants with EHR data, No. (%)** | **Coccidioidomycosis, No. (%)** | **Aspergillosis, No. (%)** | **Histoplasmosis, No. (%)** |
| *Total cohort size* | 393,601 (100) | 1,171 (100) | 687 (100) | 347 (100) |
| *Sex* | | | | |
| Female | 239,565 (60.9) | 616 (52.6) | 367 (53.4) | 185 (53.3) |
| Male | 149,871 (38.1) | 541 (46.2) | 316 (46.0) | 158 (45.5) |
| Intersex | 93 (0.02) | 0 | 0 | 0 |
| None Of These | 124 (0.03) | ≤20 | 0 | 0 |
| Skip | 3,438 (0.9) | ≤20 | ≤20 | ≤20 |
| Not Specified | 510 (0.1) | ≤20 | 0 | 0 |
| *Age* | | | | |
| 18-44 | 118,616 (30.1) | 267 (22.8) | 91 (13.2) | 34 (9.8) |
| 45-64 | 139,666 (35.5) | 433 (37.0) | 256 (37.3) | 109 (31.4) |
| 65+ | 135,319 (34.4) | 471 (40.2) | 340 (49.5) | 204 (58.8) |
| *Race* | | | | |
| White | 216,461 (55.0) | 651 (55.6) | 435 (63.3) | 265 (76.4) |
| Black | 68,534 (17.4) | 104 (8.9) | 89 (13.0) | 41 (11.8) |
| Asian | 11,690 (3.0) | 23 (2.0) | ≤20 | ≤20 |
| American Indian or Alaska Native | 6,030 (1.5) | 75 (6.4) | ≤20 | ≤20 |
| Middle Eastern or North African | 2,221 (0.6) | ≤20 | ≤20 | 0 |
| Native Hawaiian or Other Pacific Islander | 438 (0.1) | ≤20 | ≤20 | 0 |
| More Than One Population | 17,578 (4.5) | 66 (5.6) | 29 (4.2) | ≤20 |
| None Of These | 4,042 (1.0) | ≤20 | ≤20 | ≤20 |
| Skip | 4,784 (1.2) | ≤20 | ≤20 | ≤20 |
| Not Specified | 61,823 (15.7) | 217 (18.5) | 94 (13.7) | ≤20 |
| *Ethnicity* | | | | |
| Not Hispanic | 310,763 (79.0) | 859 (73.4) | 559 (81.4) | 313 (90.2) |
| Hispanic | 71,797 (18.2) | 277 (23.7) | 116 (16.9) | 21 (6.1) |
| None Of These | 4,042 (1.0) | ≤20 | ≤20 | ≤20 |
| Skip | 4,784 (1.2) | ≤20 | ≤20 | ≤20 |
| Not Specified | 2,215 (0.6) | ≤20 | ≤20 | 0 |
| *Education* | | | | |
| Never Attended | 537 (0.1) | ≤20 | ≤20 | 0 |
| One Through Four | 3,151 (0.8) | ≤20 | ≤20 | ≤20 |
| Five Through Eight | 8,550 (2.2) | 37 (3.2) | ≤20 | ≤20 |
| Nine Through Eleven | 22,537 (5.7) | 83 (7.1) | 29 (4.2) | ≤20 |
| Twelve | 74,174 (18.8) | 262 (22.4) | 135 (19.7) | 52 (15.0) |
| College One to Three | 103,000 (26.2) | 420 (35.9) | 203 (29.5) | 96 (27.7) |
| College Graduate | 89,340 (22.7) | 211 (18.0) | 148 (21.5) | 69 (19.9) |
| Advanced Degree | 83,228 (21.1) | 119 (10.2) | 140 (20.4) | 97 (28.0) |
| Skip | 6,775 (1.7) | ≤20 | ≤20 | ≤20 |
| Not Specified | 2,309 (0.6) | ≤20 | ≤20 | ≤20 |
| *Self-Reported Income* | | | | |
| Less 10K | 50,080 (12.7) | 167 (14.3) | 81 (11.8) | 31 (8.9) |
| 10K-25K | 44,533 (11.3) | 189 (16.1) | 93 (13.5) | 49 (14.1) |
| 25K-35K | 27,661 (7.0) | 71 (6.1) | 66 (9.6) | 25 (7.2) |
| 35K-50K | 31,620 (8.0) | 94 (8.0) | 50 (7.3) | 29 (8.4) |
| 50K-75K | 42,390 (10.8) | 116 (9.9) | 71 (10.3) | 39 (11.2) |
| 75K-100K | 32,951 (8.4) | 80 (6.8) | 52 (7.6) | 38 (11.0) |
| 100K-150K | 40,836 (10.4) | 79 (6.7) | 65 (9.5) | 38 (11.0) |
| 150K-200K | 19,069 (4.8) | 27 (2.3) | 26 (3.8) | ≤20 |
| More 200K | 26,112 (6.6) | 28 (2.4) | 39 (5.7) | 29 (8.4) |
| Skip | 25,744 (6.5) | 102 (8.7) | 43 (6.3) | ≤20 |
| Not Specified | 52,605 (13.4) | 218 (18.6) | 101 (14.7) | 34 (9.8) |

EHR, electronic health record; K, thousand US dollars.

**Table 4.** Host factors and antifungal treatment in the fungal infection cohorts.

| Category | Coccidioidomycosis, No. (%) | Aspergillosis, No. (%) | Histoplasmosis, No. (%) |
|---|---|---|---|
| *Conditions recorded on or before the index date* | | | |
| HIV infection | 49 (4.2) | 28 (4.1) | 28 (8.1) |
| Solid organ or hematopoietic cell transplantation | 78 (6.7) | 127 (18.5) | ≤20 |
| Hematologic malignancy | 42 (3.6) | 84 (12.2) | ≤20 |
| Other malignancy | 157 (13.4) | 132 (19.2) | 44 (12.7) |
| Neutropenia 90 days before the index date | ≤20 | 39 (5.7) | ≤20 |
| Diabetes mellitus | 460 (39.3) | 246 (35.8) | 87 (25.1) |
| COPD, emphysema, or bronchiectasis | 228 (19.5) | 336 (48.9) | 54 (15.6) |
| Asthma | 257 (21.9) | 335 (48.8) | 65 (18.7) |
| *Drugs before the index date* | | | |
| Systemic corticosteroids on at least two dates in the 60 days before[a] | 113 (9.6) | 177 (25.8) | ≤20 |
| Other immunosuppressive drugs in the 90 days before[b] | 37 (3.2) | 97 (14.1) | ≤20 |
| Immunocompromised | 269 (23.0) | 307 (44.7) | 64 (18.4) |
| Antifungal drug within 90 days before or after the index date[c] | 356 (30.4) | 307 (44.7) | 62 (17.9) |

COPD, chronic obstructive pulmonary disease; HIV, human immunodeficiency virus.

[a]Prednisone, prednisolone, methylprednisolone, dexamethasone, or hydrocortisone, excluding topical, inhaled, nasal, and rectal forms.

[b]Tacrolimus, mycophenolate, azathioprine, cyclosporine, sirolimus, adalimumab, infliximab, etanercept, or rituximab, excluding topical, and other non-systemic forms.

[c]Fluconazole, itraconazole, voriconazole, posaconazole, isavuconazole, amphotericin B, caspofungin, micafungin, or anidulafungin.

**Table 5.** Completeness of Laboratory Biomarkers Across Fungal Cohorts.

| | **Coccidioidomycosis, N (%)** | **Aspergillosis, N (%)** | **Histoplasmosis, N (%)** |
|---|---|---|---|
| **Labs /Total Cohort Size** | 1,171 (100) | 687 (100) | 347 (100) |
| Eosinophils % | 832 (71.1) | 452 (65.8) | 146 (42.1) |
| Eosinophils (absolute) | 837 (71.5) | 432 (62.9) | 134 (38.6) |
| Erythrocyte Sedimentation Rate | 212 (18.1) | 156 (22.7) | 36 (10.4) |
| Procalcitonin | ≤20 | 64 (9.3) | ≤20 |
| 1,3-β-D-glucan | ≤20 | 69 (10.0) | ≤20 |
| C-reactive protein | 297 (25.4) | 167 (24.3) | 40 (11.5) |
| Lactate dehydrogenase | 140 (12.0) | 149 (21.7) | 30 (8.6) |
| Alkaline phosphatase | 840 (71.7) | 476 (69.3) | 169 (48.7) |
| Hemoglobin | 878 (75.0) | 502 (73.1) | 190 (54.8) |
| Hematocrit | 878 (75.0) | 505 (73.5) | 195 (56.2) |
| *Coccidioides* serology or antigen | 54 (4.6) | 43 (6.3) | ≤20 |
| *Histoplasma* antigen or serology | ≤20 | 71 (10.3) | 42 (12.1) |
| *Aspergillus* galactomannan or antibody | 80 (6.8) | 255 (37.1) | ≤20 |
| Fungal culture or stain | ≤20 | 42 (6.1) | ≤20 |

**Table 6.** Unit Conformance of Laboratory Biomarkers across Fungal Cohorts.

| | Coccidioidomycosis | | | Aspergillosis | | | Histoplasmosis | | |
|---|---|---|---|---|---|---|---|---|---|
| **Labs** | **Counts** | **Unit N** | **Most Used Unit (%)** | **Counts** | **Unit N** | **Most Used Unit (%)** | **Counts** | **Unit N** | **Most Used Unit (%)** |
| Eosinophils % | 7065 | 6 | % (84.1) | 6283 | 8 | % (86.7) | 653 | 8 | % (86.8) |
| Eosinophils (absolute) | 7081 | 11 | No-match (94.5) | 6165 | 10 | 10*3/uL (40.3) | 598 | 10 | 10*3/uL (37.3) |
| Erythrocyte Sedimentation Rate | 433 | 4 | No-match (93.3) | 379 | 6 | No-match (57) | 75 | 4 | mm/h (62.7) |
| C-reactive protein | 680 | 4 | mg/L (95.1) | 566 | 5 | mg/L (56.7) | 108 | 5 | mg/L (57.4) |
| Lactate dehydrogenase | 294 | 6 | 258947003 (60.2) | 1511 | 7 | [U]/L (67) | 50 | 5 | [U]/L (70) |
| Alkaline phosphatase | 7813 | 7 | 258947003 (65.1) | 6496 | 7 | [U]/L (45.8) | 682 | 7 | [U]/L (38.9) |
| Hemoglobin | 11370 | 5 | g/dL (97.4) | 12706 | 8 | g/dL (79.2) | 1592 | 7 | g/dL (77.9) |
| Hematocrit | 11313 | 4 | % (97.5%) | 11799 | 6 | % (90.7%) | 1518 | 7 | % (91.6%) |

No-match, no matching unit concept.

## S1 Supplementary material

**Supplementary table S1** Concept IDs used to identify aspergillosis in the electronic health records, *All of Us* Research Program.

| Description | OMOP Concept ID | SNOMED Code |
|---|---|---|
| Allergic bronchopulmonary aspergillosis | 257583 | 37981002 |
| Pneumonia in aspergillosis | 261053 | 111900000 |
| Invasive aspergillosis | 36715993 | 721798004 |
| Disseminated aspergillosis | 4297886 | 7671008 |
| Subacute invasive pulmonary aspergillosis | 37203790 | 782761005 |
| Pulmonary aspergillosis | 4245409 | 6042001 |
| Aspergillosis | 434281 | 65553006 |
| Invasive pulmonary aspergillosis | 4137770 | 3214003 |

OMOP, Observational Medical Outcomes Partnership; SNOMED, Systematized Nomenclature of Medicine.

**Supplementary table S2** Concept IDs used to identify histoplasmosis in the electronic health records, *All of Us* Research Program.

| Description | OMOP Concept ID | SNOMED Code |
|---|---|---|
| *Histoplasma capsulatum* with retinitis | 378078 | 187039009 |
| Chronic pulmonary histoplasmosis | 4096917 | 26427008 |
| Pneumonia due to *Histoplasma capsulatum* | 4299862 | 38699009 |
| Ocular histoplasmosis syndrome | 4220147 | 416770009 |
| Infection by *Histoplasma capsulatum* | 437214 | 76255006 |
| Histoplasmosis | 433134 | 12962009 |
| *Histoplasma duboisii* with pericarditis | 439682 | 187050007 |
| Pulmonary African histoplasmosis | 258354 | 187052004 |
| Meningitis caused by *Histoplasma duboisii* | 37017364 | 713625005 |
| Acute pulmonary histoplasmosis | 4241693 | 58524006 |
| Pneumonia due to *Histoplasma* | 40481839 | 442094008 |
| Pulmonary histoplasmosis | 4087778 | 187054003 |
| Histoplasmosis with retinitis | 373972 | 187058000 |
| Histoplasmosis syndrome of bilateral eyes | 36713313 | 677201000119104 |
| African histoplasmosis | 434581 | 78511005 |
| *Histoplasma duboisii* with retinitis | 373963 | 187049007 |
| Histoplasmosis syndrome of left eye | 37109491 | 677211000119101 |
| Disseminated *Histoplasma capsulatum* infection | 4139537 | 425418002 |
| Histoplasmosis syndrome of right eye | 37109492 | 677221000119108 |

OMOP, Observational Medical Outcomes Partnership; SNOMED, Systematized Nomenclature of Medicine.